\documentclass[aps,pre,twocolumn,superscriptaddress,nofootinbib]{revtex4-2}

\usepackage[utf8]{inputenc}
\usepackage{amsmath,amssymb}
\usepackage{graphicx}
\usepackage{braket}
\usepackage{color}
\usepackage{hyperref}
\usepackage{lipsum}
\usepackage{soul}

\definecolor{mygreen}{RGB}{29,177,0}

\begin{document}
\title{Ruelle-Pollicott Decay of Out-of-Time-Order Correlators in Many-Body Systems}
\author{Jer\'onimo Duarte}
\affiliation{Instituto de Investigaciones F\'isicas de Mar del Plata (IFIMAR),\\ Facultad de Ciencias Exactas y Naturales,
Universidad Nacional de Mar del Plata\\ \& CONICET, Funes 3350 (7600) Mar del Plata, Argentina}

\author{Ignacio Garc\'ia-Mata}
\affiliation{Instituto de Investigaciones F\'isicas de Mar del Plata (IFIMAR),\\ Facultad de Ciencias Exactas y Naturales,
Universidad Nacional de Mar del Plata\\ \& CONICET, Funes 3350 (7600) Mar del Plata, Argentina}

\author{Diego A. Wisniacki}
\affiliation{Departamento de F\'isica ``J. J. Giambiagi'' and IFIBA, FCEyN,
Universidad de Buenos Aires, 1428 Buenos Aires, Argentina}

\date{\today}

\begin{abstract}
The out-of-time-order correlator (OTOC) quantifies information scrambling in quantum systems and serves as a key diagnostic of quantum chaos. In one-body systems with a classical counterpart, the relaxation of the OTOC is governed by Ruelle-Pollicott resonances. For many-body systems lacking a semiclassical limit, recent studies have identified an analogous role played by the Liouvillian spectrum of weakly open extensions of the dynamics, where the slowest decay rate -- the Liouvillian gap -- encodes relaxation. Here we study the kicked Ising spin chain and show that the long-time exponential decay of the OTOC in the isolated system occurs at a rate equal to twice this intrinsic gap. This correspondence is demonstrated across parameter regions exhibiting distinct level spacing statistics, indicating that the Liouvillian spectrum provides a robust framework for characterizing relaxation and irreversibility in closed many-body quantum systems.
\end{abstract}
\maketitle
%%%
\section{Introduction}
Understanding relaxation dynamics and the emergence of effective irreversibility in quantum many-body systems is a central problem at the interface of quantum chaos, statistical mechanics, and open-system theory. In classical chaotic dynamics, the intermediate time decay of correlations is governed by the Ruelle-Pollicott (RP) resonances \cite{ruelle1986locating,ruelle1986PRL,pollicott1985rate} -- isolated singularities of the analytically continued spectrum of the Frobenius-Perron operator -- which generate effective irreversibility despite the underlying deterministic and reversible motion. In quantum one-body systems with a classical limit, an analogous correspondence is well established through spectral and operator-truncation approaches \cite{manderfeld2001classical,khodas2000relaxation,fishman2002relaxation} or through weakly dissipative formulations \cite{nonnenmacher2003spectral,GarciaMata2004,garciama2005}.

In contrast, identifying the quantum analog of RP resonances in generic many-body systems remains an open question. Several works  \cite{prosen2002ruelle,prosen2004ruelle,prosen2007chaos} have introduced coarse-grained formulations of the operator space, revealing isolated eigenvalues inside the unit disk that can be interpreted as quantum RP resonances. While conceptually appealing, this approach becomes impractical for large systems due to the exponential growth of the operator space and the difficulty of defining an appropriate coarse-graining. A more recent development exploits translational invariance and quasi-momentum decomposition \cite{znidaric2024momentum}, enabling the identification of resonances within each symmetry sector.

An alternative line of research focuses on the Liouvillian spectrum of weakly open quantum systems \cite{mori2024liouvillian}. When a system is coupled to an environment, its relaxation dynamics are governed by the eigenvalues of the corresponding Lindblad superoperator, with the slowest nonzero decay rate defining the Liouvillian gap. In exactly solvable models \cite{jacoby2025spectral,zhang2024thermalization,yoshimura2025irreversibility}, this gap has been shown to coincide with the leading RP resonance of the corresponding isolated dynamics. 

The magnitude of the leading RP resonance determines the approach to equilibrium of dynamical observables, including the out-of-time-order correlator (OTOC). The OTOC measures the spread of local perturbations and serves as a sensitive probe of information scrambling in quantum systems. Although the butterfly effect in quantum chaos is not universal \cite{garciamata2023scholar}, the relaxation of the OTOC at late times provides a robust signature of underlying dynamics. 
In one-body systems, the crossover from integrability to chaos can be characterized through the OTOC decay rate \cite{garciamata2018PRL}, a result that extends to systems with mixed classical dynamics \cite{notenson2023classical}. 
More recently, a large-deviation theory of freeness has been proposed to describe the decay of higher-order OTOCs, introducing a hierarchy of characteristic time scales and mono- or multifractal behavior \cite{vallini2024freeness}. Within this framework, a possible interpretation of these behaviors in terms of Ruelle-Pollicott resonances has been suggested, but remains an open question. The connection was  further generalized for higher-order OTOCS in minimal quantum circuits and Floquet systems, where exact results were obtained. In particular, for chaotic many-body models the OTOC decay rate is governed by the subleading eigenvalue of the corresponding quantum channel\cite{fritzsch2025free,fritzsch2025freecumulants}.\\
Here we extend this connection to many-body systems. We study the kicked Ising spin chain, a paradigmatic model widely used in the study of quantum dynamics. Specifically, we analyze three complementary quantities: (i) the Liouvillian gap $\bar{g}$ extracted from a weakly dissipative extension of the dynamics, (ii) standard finite-size diagnostics based on spectral statistics and eigenstate delocalization, and (iii) the intermediate-time exponential decay rate $\alpha$ of the OTOC in the isolated chain. We show that $\alpha \approx 2\bar{g}$ across parameter regions displaying markedly different spectral and dynamical behavior. This establishes the Liouvillian gap as a robust indicator of long-time relaxation and operator spreading, independently of the specific finite-size statistical regime.
%%%%%%%%%%%%%%%%%%%%%%%%%%%%%%%%
\section{Many-body System: Kicked Ising Spin Chain}\label{sec:system}
%%%%%%%%%%%%%%%%
To investigate the interplay between relaxation dynamics, dissipation, and operator spreading in a many-body setting, we consider the kicked Ising spin chain~\cite{Prosen2000,Prosen2002PRE}, a paradigmatic model of Floquet many-body dynamics. This system is particularly well suited to our study because its parameter space spans regions with markedly different spectral and dynamical properties, allowing us to test the robustness of our results across distinct regimes.

The model is defined by the time-periodic Hamiltonian
\begin{equation}
    \hat{H}(t) = \hat{H}_{\mathrm{free}} + \hat{H}_{\mathrm{kick}} \, \tau \sum\limits_{n=-\infty}^{\infty} \delta(t - n\tau),
\end{equation}
where the free and kick components take the form
\begin{equation}
\begin{cases}
\hat{H}_{\mathrm{free}} = - J \sum\limits_{i=0}^{L-2} \hat{\sigma}_i^z \hat{\sigma}_{i+1}^z - h_z \sum\limits_{i=0}^{L-1} \hat{\sigma}_i^z, \\[10pt]
\hat{H}_{\mathrm{kick}} = - h_x \sum\limits_{i=0}^{L-1} \hat{\sigma}_i^x.
\end{cases}
\end{equation}
Here, $\hat{\sigma}_i^{\alpha}$  (\( \alpha = x, z \)) denote Pauli operators acting on site \( i \), \( J \) is the nearest-neighbor coupling, and \( h_x \), \( h_z \) are transverse and longitudinal magnetic fields, respectively. The dynamics is stroboscopic with period \( \tau \), and the evolution over one period is governed by the Floquet operator
\begin{equation}
    \hat{U}_F = \hat{U}_{\mathrm{free}} \hat{U}_{\mathrm{kick}},
\end{equation}
where
\begin{equation}
\begin{cases}
    \hat{U}_{\mathrm{free}} = \exp(-i \hat{H}_{\mathrm{free}} \tau), \\[10pt]
    \hat{U}_{\mathrm{kick}} = \exp(-i \hat{H}_{\mathrm{kick}} \tau).
\end{cases}
\end{equation}

This Floquet representation provides a convenient description of the unitary stroboscopic dynamics, which later serves as a natural starting point for incorporating weak dissipation at the level of the evolution superoperator.

Throughout this work we set $J = h_z = \tau = 1$ and vary the transverse field $h_x$, which controls the degree of chaos. 
These values ensure nontrivial dynamics while reducing the parameter space to a single effective control variable. 
All simulations are performed for system sizes up to $L = 12$ spins.

To understand the structure of the spectrum, it is essential to discuss the symmetries present in the model. Since we work with a Floquet operator, the relevant spectrum consists of quasienergies. Throughout this study we impose open boundary conditions (OBC). The system possesses an external reflection symmetry \( \hat{R} \), which acts as
\[
\hat{R}\ket{m_1, m_2, \dots, m_{L-2}, m_{L-1}} = \ket{m_{L-1}, m_{L-2}, \dots, m_2, m_1},
\]
where \(\ket{m_1, m_2, \dots, m_{L-1}}\) denotes a computational-basis state.
Because $[\hat{U}_F, \hat{R}] = 0$, the Hilbert space decomposes into two invariant subspaces corresponding to the eigenvalues $+1$ and $-1$ of $\hat{R}$, which we refer to as the even and odd parity sectors. 
This symmetry decomposition allows us to analyze each sector independently, avoiding spectral mixing and enabling a cleaner identification of chaotic behavior and Liouvillian relaxation modes.
%%%%%%%%
\section{Quantum Chaos Diagnostics}

To characterize the spectral and dynamical properties of the kicked Ising spin chain, we employ two complementary diagnostics: spectral statistics and eigenstate delocalization. These indicators are commonly used to assess chaotic versus integrable-like behavior in finite quantum systems and serve to organize the different parameter regions considered in our analysis, providing context for the behavior of the OTOC decay and its relation to the Liouvillian gap.

A well-established signature of quantum chaos is the statistical distribution of level spacings~\cite{bohigas1984characterization}. Chaotic systems exhibit Wigner-Dyson distributions predicted by random matrix theory (RMT), with the specific ensemble---in our case, the circular orthogonal ensemble (COE)---determined by the symmetries of the system~\cite{mehta2004random}. In contrast, integrable systems display Poisson statistics reflecting the presence of extensive conserved quantities~\cite{berrytabor}.

Rather than computing the full level-spacing distribution, we employ the ratio of adjacent spacings~\cite{oganesyan2007localization}, which provides a practical and robust diagnostic. For the Floquet operator satisfying
\begin{equation}
\hat{U}_F \ket{\psi_n} = e^{i\varphi_n} \ket{\psi_n}, \quad n=1,2,\dots,D,
\end{equation}
where $\varphi_n$ are the quasienergies, $|\psi_n\rangle$ are the eigenstates, and $D$ is the Hilbert space dimension, we compute
\begin{equation}
    r_n = \frac{\min(\delta_n, \delta_{n-1})}{\max(\delta_n, \delta_{n-1})}, \qquad \delta_n = \varphi_{n+1} - \varphi_n.
\end{equation}
The mean value $\langle r \rangle$ exhibits two distinct limits: $\langle r \rangle_{\text{COE}} \approx 0.5307$ for COE statistics and $\langle r \rangle_P \approx 0.3863$ for Poisson statistics~\cite{atas2013distribution,Lee1985RMP}. To interpolate between these limits, we define the normalized parameter
\begin{equation}
\eta = \frac{\langle r \rangle - \langle r \rangle_P}{\langle r \rangle_{\text{COE}} - \langle r \rangle_P},
\label{eq:eta}
\end{equation}
which provides a continuous measure ranging from $\eta \to 0$ in the integrable regime to $\eta \to 1$ in the fully chaotic regime.

A complementary diagnostic is provided by eigenstate delocalization, quantified through the participation ratio (PR)~\cite{wegner1980inverse}. Consider an eigenstate $\ket{\psi_i}$ expanded in a reference basis $\{\ket{\phi_j}\}_{j=0}^{D-1}$ as $\ket{\psi_i} = \sum_j a_{ij} \ket{\phi_j}$. The participation ratio is defined as 
\begin{equation}
\xi_E(i) = \left[\sum_{j=0}^{D-1} |a_{ij}|^4\right]^{-1}.
\label{eq:PR}
\end{equation}
This quantity measures how extended an eigenstate is in the chosen basis. Small values indicate localization, while high values signal delocalization. For chaotic systems consistent with RMT, the coefficients \( |a_{ij}|^2 \) behave as independent random variables, leading to typical values \( \xi_E^{\mathrm{deloc}} \approx D/3 \), where \( D \) is the Hilbert space dimension \cite{AmJPhys.80.246, PhysRep.276.85}. In our numerical analysis, we compute the average of \( \xi_E(i) \) over all eigenstates.

These spectral and eigenstate-based diagnostics provide a quantitative characterization of the \st{finite-size} dynamical properties of the system. We analyze their behavior as a function of the transverse field strength \( h_x \), which controls the kick strength in the Floquet dynamics. To avoid spectral mixing between symmetry sectors, all analyses are performed within the even-parity subspace, ensuring a consistent comparison across parameter values. This characterization provides a reference framework for the subsequent analysis of the Liouvillian gap and the OTOC decay.
%%%%%%%%%%%%%%%%%%%%%%%%%%%%%%%%%%%%%%%%%
\section{Out-of-Time-Order Correlator}
%%%%%%%%%%%%%%%%%%%%%%%%%%%%%%%%%%%%%%%%%%%
The out-of-time-order correlator (OTOC) quantifies the scrambling of quantum information and is defined as the thermal expectation value of the squared commutator between two operators evaluated at different times:
\begin{equation}
    C(t) = \left\langle \left[ \hat{W}(t), \hat{V} \right] \left[ \hat{W}(t), \hat{V} \right]^\dagger \right\rangle,
\end{equation}
where $\hat{W}(t)$ is the Heisenberg picture evolution of $\hat{W}$. Expectation values are taken over the infinite-temperature ensemble,
$\langle \cdot \rangle = \mathrm{Tr}(\cdot)/D$, with $D$ the Hilbert-space dimension.
This ensemble is appropriate for Floquet systems or Hamiltonians with bounded spectra, where microcanonical and infinite-temperature averages coincide in the thermodynamic limit.

Originally introduced in studies of superconductivity~\cite{larkin1969quasiclassical}, the OTOC has become a central tool for characterizing quantum chaos and information scrambling in many-body systems~\cite{Xu2024scrambling}. The OTOC gained attention when upper bound for the exponential growth was found \cite{maldacena2016bound}, setting the bar of fast scramblers like black holes\cite{shenker2013black} and strongly correlated fermionic systems\cite{sachdev1993,Kitaev,maldacena2016remarks}.
In systems with a classical limit, the growth rate coincides with the classical Lyapunov exponent~\cite{rozenbaum2017PRL,garciamata2018PRL,Jalabert2018,operScramblingQC}.
At later times, quantum interference leads to saturation around a constant plateau with residual fluctuations that depend on the underlying dynamics~\cite{fortesgauging}.

We focus on local Hermitian and unitary operators $\hat{W}$ and $\hat{V}$ acting on distinct sites of the spin chain. A convenient choice is given by Pauli operators $\hat{\sigma}^\mu_i$ ($\mu = x,y,z$) acting on site $i$.
At infinite temperature, the OTOC takes the explicit form
\begin{equation}
\begin{aligned}
    C^{\mu\nu}(l,t) &= \frac{1}{2} \left\langle \left[ \hat{\sigma}_0^\mu(t), \hat{\sigma}_l^\nu \right]^2 \right\rangle \\
    &= 1 - \frac{1}{D} \, \mathrm{Re} \left\{ \mathrm{Tr} \left[ \hat{\sigma}_0^\mu(t) \hat{\sigma}_l^\nu \hat{\sigma}_0^\mu(t) \hat{\sigma}_l^\nu \right] \right\}.
\end{aligned}
\end{equation}
We define 
\begin{equation}
    O_1^{\mu\nu}(l,t) = \frac{1}{D} \, \mathrm{Re} \left\{ \mathrm{Tr} \left[ \hat{\sigma}_0^\mu(t) \hat{\sigma}_l^\nu \hat{\sigma}_0^\mu(t) \hat{\sigma}_l^\nu \right] \right\},
\end{equation}
so that $C^{\mu\nu}(l,t) = 1 - O^{\mu\nu}_1(l,t)$.
In what follows we focus on the case $\mu=\nu=z$, denoted $C^{zz}(l,t)$, which probes the scrambling and relaxation of local $\hat{\sigma}^z$ operators.

At intermediate times, 
the approach of $C(t)$ to its saturation is well described by an exponential decay of $O_1(t)$, which can occur even in systems that are not fully chaotic~\cite{notenson2023classical}. For one-body systems with a well-defined classical limit, such as quantum maps, this decay has been shown to be governed by classical Ruelle-Pollicott resonances~\cite{garciamata2018PRL}, which are extracted from a coarse-grained propagator acting on phase-space distributions.

In contrast, the kicked Ising spin chain considered here is a quantum many-body system that does not admit a classical phase-space description. As a consequence, the mechanisms controlling the long-time decay of the OTOC must be characterized using intrinsically quantum tools, which will be introduced in the following sections.

\begin{figure}[h]
    \centering
    \includegraphics[width=0.95\linewidth]{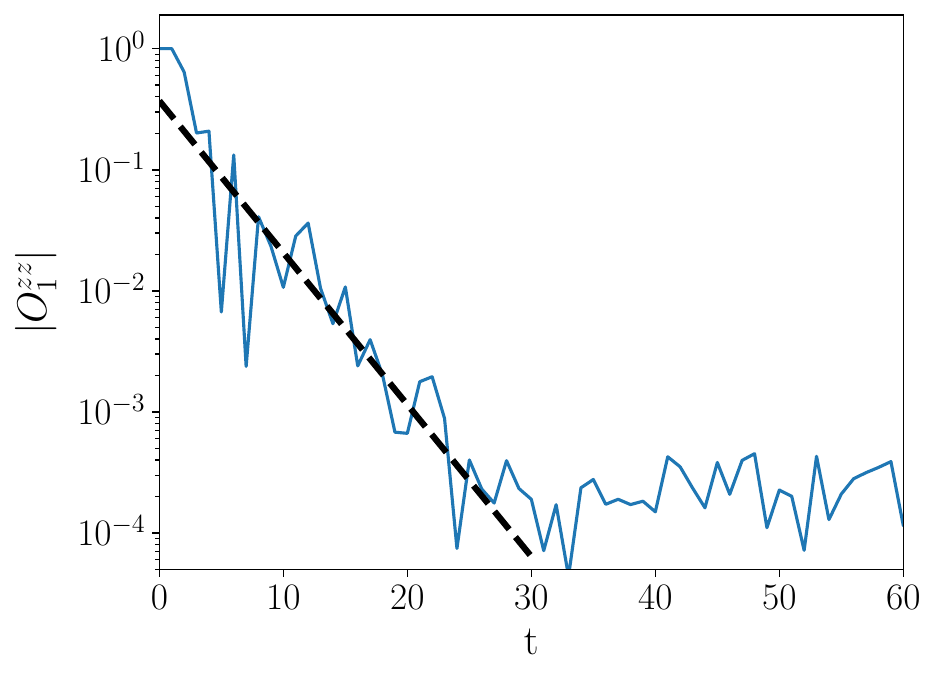}
    \caption{The absolute value of \( O_1^{zz}(1,t) \) for a transverse field \( h_x \) in the chaotic regime. The data corresponds to a spin chain of size \( L = 12 \) and $h_x=0.8168$. The blue line with circles represents \( O_1^{zz}(1,t) \), while the dashed line is the exponential fit to the intermediate time behavior of \( |O_1^{zz}(1,t)| \), with a decay exponent $\alpha =$ \( 0.2888 \).}
    \label{fig:otoc_decay}
\end{figure}

Figure~\ref{fig:otoc_decay} illustrates the time evolution of
$|O^{zz}_1(l,t)|$ for $l=1$ and a transverse field $h_x$ deep in the chaotic regime. The data, obtained for $L=12$, display a clear exponential decay at intermediate times
. An exponential fit of the form
$|O^{zz}_1(1,t)| \propto e^{-\alpha t}$
yields the decay rate $\alpha$, which we compare below with twice the Liouvillian relaxation rate extracted from the weakly dissipative extension of the model.

The decay exponent $\alpha$ is extracted from the exponential relaxation of the OTOC in the intermediate-time window defined above. Importantly, this exponent is determined from a time regime in which the dynamics is no longer sensitive to the specific initial condition. In this window, transient effects have decayed while saturation has not yet set in, and the OTOC displays a well-defined exponential behavior.

This allows for a reliable and unambiguous extraction of the decay rate $\alpha$. Since the focus of this work is on the exponential relaxation preceding saturation, all reported values of $\alpha$ are consistently obtained from this intermediate-time regime.

%%%%%%%%%%%%%%%%%%%%%%%%%%%%%%%%%%%%%%%%%%%%%%
\section{Liouvillian Gap}
%%%%%%%%%%%%%%%%%%%%%%%%%%%%%%%%%%%%%%%%%%%%%%%
To investigate the relation between the decay rate discussed in the previous section and the spectral properties of the system, we follow Ref.~\cite{mori2024liouvillian}. In this approach, the system is extended to a weakly open setting described by a Liouvillian superoperator
\begin{equation}
\mathcal{L} = -i[H,\ \cdot\ ] + \gamma \mathcal{D},
\label{eq:Lindblad}
\end{equation}
where $\mathcal{D}$ is a dissipator and $\gamma$ the dissipation strength.
Building on the kicked Ising model defined above, 
we introduce bulk dephasing and study the corresponding relaxation dynamics through the time-periodic Lindblad master equation
\begin{eqnarray}
\frac{d\rho}{dt}
&=& -\,i\,[\hat H(t),\rho]
+ \gamma \sum_{i=0}^{L-1}
\Bigl(\sigma_i^z\,\rho\,\sigma_i^z \;-\;\tfrac12\{\sigma_i^z\sigma_i^z,\rho\}\Bigr)\nonumber\\
&\equiv& \mathcal{L}(t)\,\rho,
\end{eqnarray}
where \(\gamma>0\) controls the dephasing strength and \(\hat H(t)=\hat H(t+\tau)\) is the time-dependent Hamiltonian.  

The Liouvillian superoperator $\mathcal{L}(t)$ has a complex spectrum  ${\lambda_\alpha}$ with negative real parts.
For time-independent generators ($\mathcal{L}$) there is one zero mode ($\lambda_0=0$) corresponding to the steady state, plus decaying modes with $\mathrm{Re}\,\lambda_\alpha<0$.
The Liouvillian gap
\begin{equation}
g = - \max_{\alpha\neq0} \mathrm{Re}\,\lambda_\alpha
\label{eq:L_gap_def}
\end{equation}
characterizes the slowest non-trivial relaxation rate.  
In periodically driven systems it is convenient to define the Floquet map
\begin{equation}
\mathcal{U}_F \;=\;\mathcal{T}\,e^{\int_0^\tau\!\mathcal{L}(t)\,dt} 
\label{eq:Floquet_map}
\end{equation}
where $\mathcal{T}$ denotes time ordering.
The eigenvalues of $\mathcal{U}_F$ are $e^{\lambda_\alpha \tau}$, allowing one to extract $g$ equivalently from the Floquet-Liouvillian spectrum.

\begin{figure}[h]
    \includegraphics[width=0.95\linewidth]{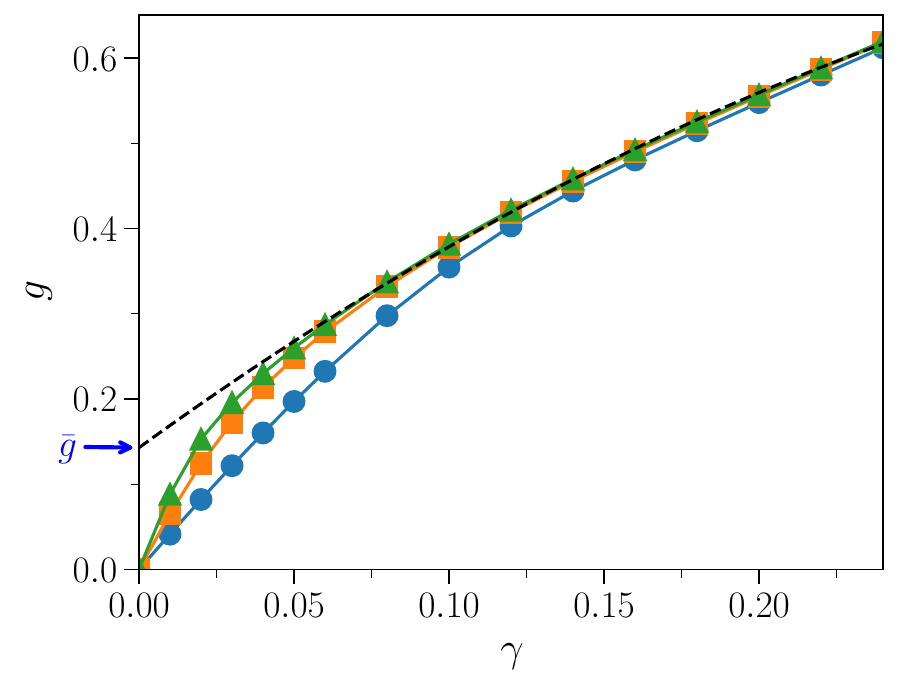} %%{hx=0.8168.pdf}
       \caption{Liouvillian gap \( g(\gamma) \) as a function of the dissipation strength \( \gamma \) for system sizes \( L = 6 \) (blue circles), \( L = 8 \) (orange squares), and \( L = 10 \) (green triangles), and $h_x = 0.8168$. The dashed line indicates quadratic fit. For \( L = 10 \), the extrapolated value of $\bar{g}$  is $0.1429$}
    \label{fig:g_vs_gamma}
\end{figure}

In chaotic many-body systems, it has been shown that taking the limits $L \to\infty$ first and $\gamma \to 0$, 
the gap converges to a nonzero limit
\begin{equation}
\bar{g} =
\lim_{\gamma\to0}
\lim_{L\to\infty} g,
\label{eq:gbar_def}
\end{equation}
which coincides with the real part of the leading Ruelle-Pollicott resonance of the corresponding isolated chain \cite{mori2024liouvillian}. 

To extract $\bar{g}$ in the weak dissipation regime, we compute \( g(\gamma) \) for several small values of \(\gamma\) and perform a quadratic fit in \(\gamma\). Following Eq.~(\ref{eq:gbar_def}), the fit is applied to the largest available system size. The eigenvalues of the Floquet-Liouvillian spectrum, from which \( g(\gamma) \) is obtained, are computed using the Arnoldi-Lindblad method described in the Appendix. This procedure provides the extrapolated value $\bar{g} \equiv \lim_{\gamma\to0} \lim_{L\to\infty} g(\gamma)$, characterizing the intrinsic relaxation of the isolated chain. 

Figure \ref{fig:g_vs_gamma} shows the Liouvillian gap $g(\gamma)$ as a function of the dissipation strength for system sizes $L=6,8,10$ at $h_x=0.8168$.
Dashed lines indicate quadratic fits, and for $L=10$ the extrapolated value $\bar{g}=0.1429$.
Remarkably, the corresponding OTOC decay rate obtained in the isolated chain satisfies $\alpha \simeq 2\bar{g}$, demonstrating that the Liouvillian relaxation rate of the weakly open system predicts the intermediate time OTOC decay in the closed dynamics.

\noindent\textit{Parity-resolved analysis}.\
We further analyze the Liouvillian spectrum by resolving it into the even and odd parity subspaces. As system parameters are varied, particularly across crossover regions between integrable-like and chaotic-like behavior, the leading gaps associated with these two subspaces may cross as functions of the dissipation strength. Consequently, the smallest nonzero Liouvillian eigenvalue—and therefore the global Liouvillian gap—can originate from either the even or the odd subspace, depending on the parameter regime. This results in an effective switching of the symmetry sector that hosts the slowest decaying mode when tracking only the leading relaxation rate. Importantly, this behavior reflects finite-size effects and does not signal any mixing or degeneracy between symmetry sectors. Rather, it indicates that different symmetry sectors can control the long-time relaxation in different parameter regions.

The Arnoldi-Lindblad approach employed here consistently captures the dominant Liouvillian eigenvalue governing relaxation, independently of the symmetry sector from which it originates. This allows us to resolve the behavior of the Liouvillian gap across a wide range of parameters, including finite-size crossover regimes where multiple symmetry sectors may compete in setting the slowest decay rate.

To carry out this analysis, we explicitly constructed both the Lindbladian superoperator and the spatial inversion superoperator. By diagonalizing the latter, we obtained its eigenvalues and eigenvectors, which allowed us to separate the Hilbert space into parity-even and parity-odd subspaces. Since the Lindbladian commutes with the spatial inversion operator, it can be represented in this eigenbasis as a block-diagonal matrix, with one block corresponding to each parity sector. Restricting the dynamics to each block separately, we extracted the eigenvalues of the Lindbladian within each subspace, and thereby determined the corresponding parity-resolved Liouvillian gaps.

\begin{figure}[h]
    \centering
    \begin{minipage}{\linewidth}
        \centering
        \includegraphics[width=\linewidth]{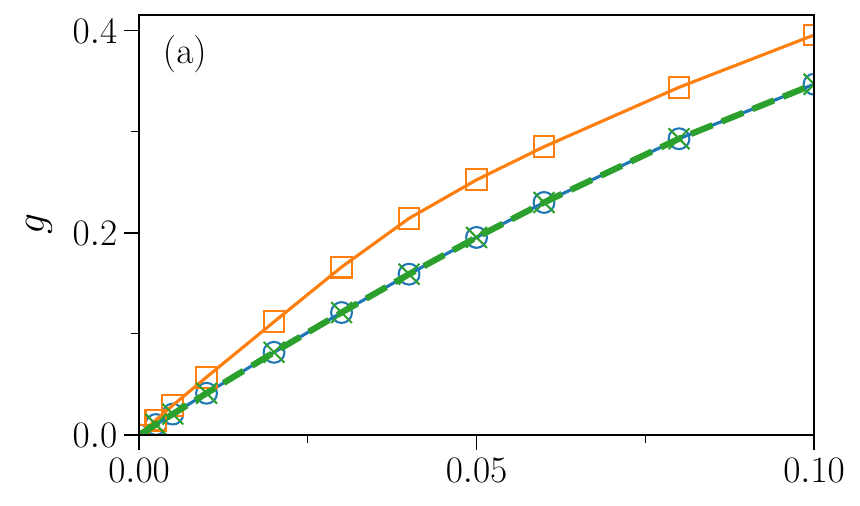} %{hx=0.7854.pdf}
        \label{fig:g_parity_a}
    \end{minipage}
    \vspace{-10pt}
    \begin{minipage}{\linewidth}
        \centering
        \includegraphics[width=\linewidth]{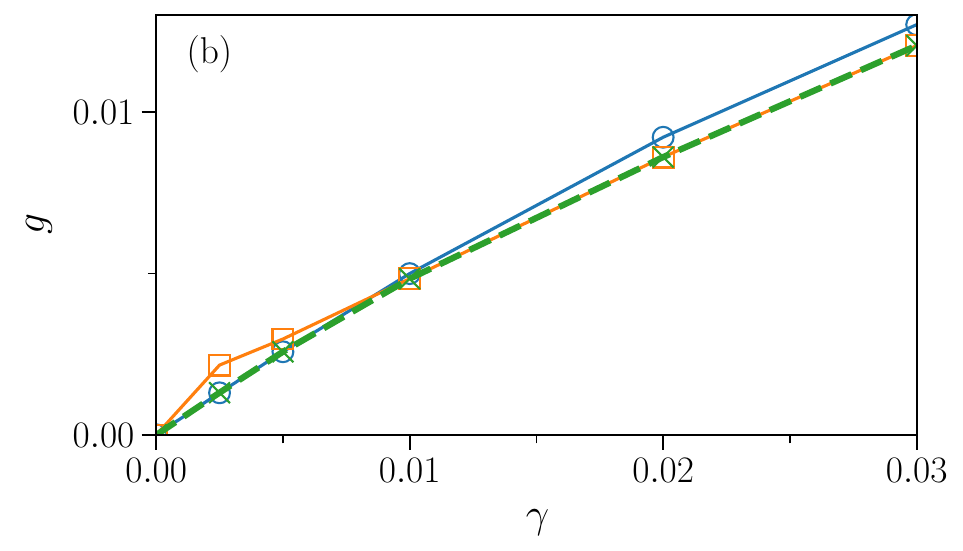} %{hx=0.2749.pdf}
        \label{fig:g_parity_b}
    \end{minipage}
    \caption{ Comparison of Liouvillian gaps for two values of the transverse field \( h_x \).
    (a) \( h_x = 0.7854 \). The parity-even subspace is shown in blue empty circles, the parity-odd subspace in orange empty squares, and the global gap obtained via the Arnoldi-Lindblad method in green crosses.
    (b) \( h_x = 0.2749 \). The parity-
    even subspace is shown in blue empty circles, the parity-
    odd subspace in orange empty squares, and the global gap obtained via the Arnoldi-Lindblad method in green crosses. }
    \label{fig:two_panels}
\end{figure}

Figure \ref{fig:two_panels} compares the Liouvillian gaps computed within each parity subspace -- parity-even (blue empty circles) and parity-odd (orange empty squares)-with the global gap obtained via the Arnoldi-Lindblad method (green crosses). Solid lines serve as guides to the eye for the subspace gaps.
As the dissipation parameter is varied, the dominant eigenvalue can switch between subspaces: in certain regions the even sector exhibits the smallest eigenvalue, while in others the odd sector becomes dominant. The Arnoldi method naturally captures this transition, since it always selects the eigenvalue with the smallest real part across all subspaces. Moreover, as the system size increases, the spectra of the two parity sectors become increasingly separated, making such crossings less probable. This behavior highlights that the Arnoldi approach consistently identifies the slowest decaying mode of the global dynamics, regardless of the underlying sectoral structure.

The case analyzed in Fig.~\ref{fig:two_panels}(a) corresponds to a parameter region where the Liouvillian gaps associated with the two parity subspaces are clearly separated over the range of dissipation strengths considered. In this regime, a well-defined hierarchy of decay rates emerges, and the slowest relaxation mode consistently belongs to a single symmetry sector. As a consequence, no crossings between the leading gaps of different parity sectors are observed. In contrast, Fig.~\ref{fig:two_panels}(b) illustrates a parameter region displaying integrable-like finite-size behavior, where the Liouvillian gap curves associated with the two parity subspaces partially overlap and crossings occur as the dissipation strength is varied. In this case, the slowest decaying mode can originate from either subspace, leading to an effective switching of the sector that controls the long-time relaxation.

%%%%%%%%%%%%%%%%
\section{Relation between OTOC decay, the Liouvillian gap and chaos parameters}
%%%%%%%%%%%%%%
In this section, we bring together the chaos indicators, OTOC decay exponents, and the Liouvillian gap - into a single comparative framework. Previously, we have shown how to compute the normalized level-spacing ratio \(\eta\) and participation ratio \(\bar{\xi}_E\) across different transverse fields \(h_x\), how to extract the intermediate time exponential decay exponent \(\alpha\) of the local correlator \(O_1^{zz}(l=1,t)\), and how to obtain the Liouvillian gap \(\bar{g}\) via quadratic extrapolation in the weak-dissipation limit for \(L=10\).

\begin{figure}[h]
    \centering
    \begin{minipage}{\linewidth}
        \centering
        \includegraphics[width=\linewidth]{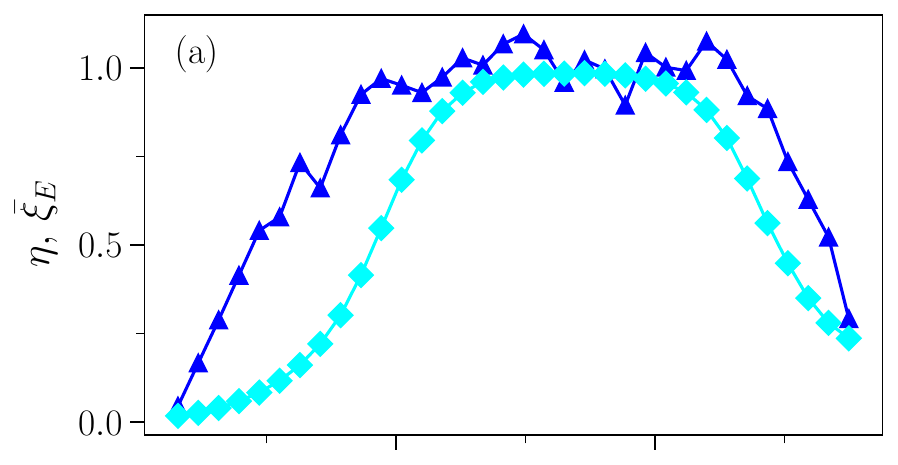} 
    \end{minipage}
    \vspace{-6pt} %  
    
    \begin{minipage}{\linewidth}
        \centering
        \includegraphics[width=\linewidth]{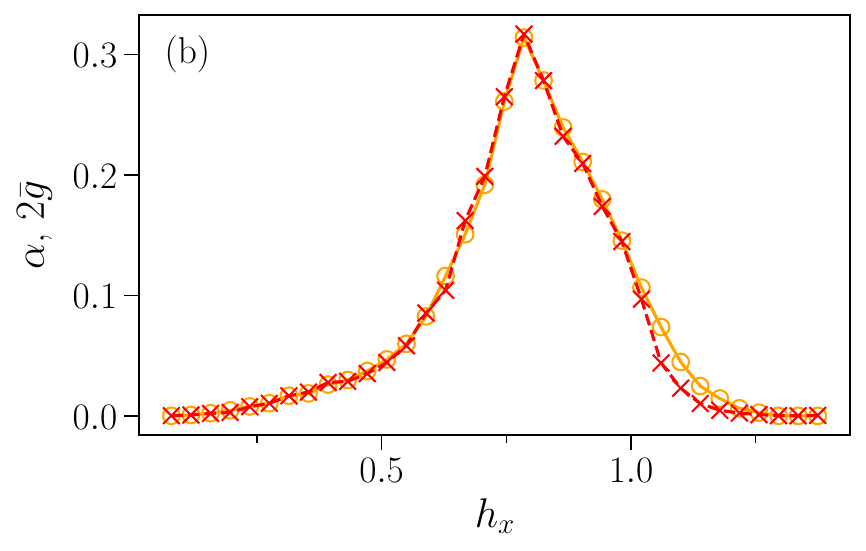}
    \end{minipage}
    \caption{(a) Chaos indicators \(\eta\) and \(\bar{\xi}_E\) as functions of the transverse field \(h_x\), computed from the Floquet spectrum in the parity-even subspace for system size \(L=12\). Blue upward triangles denote \(\eta\) and cyan diamonds denote \(\bar{\xi}_E\). (b) Comparison between the OTOC decay exponent \(\alpha\) (red crosses) and twice the Liouvillian gap \(2\bar{g}\) (orange empty circles) as functions of \(h_x\). The Liouvillian gap \(\bar{g}\) is extrapolated for \(L=10\), while \(\alpha\) is obtained from the isolated chain with \(L=12\). The two panels together illustrate how both dynamical and spectral quantities capture the transition between integrable and chaotic regimes.}
    \label{fig:conjunto}
\end{figure}

Figure~\ref{fig:conjunto}(a) displays the chaos indicators \(\eta\) (blue upward triangles) and \(\bar{\xi}_E\) (cyan diamonds) as functions of the transverse field \(h_x\). Both quantities are computed from the unitary Floquet spectrum at system size \(L=12\), and—owing to the symmetries of the model—are analyzed within the parity-even subspace defined with respect to the external reflection operator. These indicators capture a crossover between integrable-like and chaotic-like spectral and eigenstate properties as the control parameter \(h_x\) is varied.

Figure~\ref{fig:conjunto}(b) shows the comparison between the OTOC decay exponent \(\alpha\) (red crosses) and twice the Liouvillian gap \(2\bar{g}\) (orange empty circles) as functions of \(h_x\). For each value of \(h_x\), \(\alpha\) is obtained by fitting \(|O_1^{zz}(1,t)|\) to an exponential decay at intermediate times in the isolated chain of size \(L=12\), while \(\bar{g}\) is extrapolated from \(g(\gamma)\) for \(L=10\).

The comparison shown in Fig.~\ref{fig:conjunto}(b) is robust against the parity-resolved structure of the Liouvillian spectrum. As discussed above, depending on the parameter regime, the slowest decaying mode can originate from either the even or the odd parity subspace. The Arnoldi-Lindblad approach naturally captures this behavior by tracking the smallest nonzero Liouvillian eigenvalue, independently of the symmetry sector from which it arises. As a result, the extracted gap $\bar{g}$ consistently reflects the dominant relaxation rate that governs the long-time decay of the OTOC.

Standard spectral indicators of quantum chaos are included here as a complementary reference. While these quantities are known to be sensitive to finite-size effects, they provide an independent characterization of how the system parameters explored span regimes with qualitatively different finite-size dynamical behavior. Importantly, the agreement between the OTOC decay rate and twice the Liouvillian gap is found to persist across the entire parameter range, demonstrating the robustness of the Liouvillian-OTOC correspondence with respect to variations in the underlying finite-size dynamics.

Overall, this figure highlights that the OTOC decay exponent remains approximately equal to twice the Liouvillian gap for all \(h_x\) values where scrambling-induced relaxation persists. For \(h_x > 1.00\), the correspondence becomes less pronounced, likely due to finite-size effects-larger system sizes would be required to achieve quantitative convergence of both \(\alpha\) and \(\bar{g}\) in this regime. Nevertheless, even at these modest sizes, a clear qualitative and quantitative agreement is observed, underscoring the robustness of the relationship between the OTOC decay rate and the Liouvillian gap.

It is worth noting that the decay exponent $\alpha$ and the Liouvillian gap $\bar{g}$ display a correlated dependence on the system parameters. A similar qualitative (and quantitative) behavior of the OTOC decay was previously reported for quantum maps in Ref.~\cite{notenson2023classical}.

\section{Conclusions}
In this work we have established a direct connection between the intermediate-time exponential decay of out-of-time-order correlators (OTOCs) and the spectral properties of the Liouvillian in a paradigmatic many-body quantum system, the kicked Ising spin chain.
Our central finding is that the intermediate time exponential decay rate of the OTOC for generic, symmetry-unresolved operators equals twice the Liouvillian gap of the corresponding weakly open extension of the dynamics.
While this relation was previously established in systems with a well-defined semiclassical limit, our results show that it also holds in a genuine many-body quantum system when tuning parameters across different dynamical regimes at finite system size, ranging from integrable-like to chaotic behavior. This shows that the Liouvillian spectrum provides a unified framework to characterize relaxation and irreversibility in quantum many-body dynamics.

A detailed parity-resolved analysis confirms that the intrinsic Liouvillian gap extracted from the Arnoldi-Lindblad method reproduces the global relaxation rate of the system, even when the dominant decay mode shifts between symmetry sectors.
This robustness indicates that both the OTOC and the Liouvillian gap respond coherently to changes in the dynamical behavior of the system at finite size, providing consistent relaxation signatures across parameter regimes ranging from integrable-like to chaotic.

The practicality of the Arnoldi-Lindblad approach lies in its ability to access the smallest Liouvillian eigenvalues without constructing the full superoperator, enabling efficient computation of the Liouvillian gap and its extrapolation to the weak-dissipation limit.
In contrast to the momentum-resolved analysis of Ref.~\cite{znidaric2024momentum} for systems with periodic boundary conditions, our framework captures the global relaxation properties of the chain under open boundaries, providing a complementary perspective that links spectral relaxation modes directly to information scrambling.

Overall, our results establish a quantitative bridge between unitary and dissipative descriptions of quantum relaxation across a wide range of parameters.
This connection suggests that Liouvillian spectroscopy may serve as a general diagnostic tool for intermediate-time dynamics and operator spreading in isolated quantum systems.

\section*{ACKNOWLEDGMENTS}
We acknowledge support from Argentinian Agencia I$+$D$+$i (Grants No. PICT-2020-SERIEA-00740 and PICT-2020-SERIEA-01082). I.G-M received support from the French-Argentinian International Research Project \textit{Complex Quantum Systems} (COQSYS), funded by CNRS.
D.A.W.  received support from CONICET (Grant No. PIP 11220200100568CO), UBACyT (Grant No. 20020220300049BA) and PREI-UNAM.

\appendix

\section{Arnoldi-Lindblad time evolution}
\label{ap:Arnoldi}
To compute the Liouvillian spectrum efficiently, we employ the {Arnoldi--Lindblad approach} of Ref.~\cite{minganti2021arnoldi}. This method allows one to access the smallest-magnitude eigenvalues of the Floquet-Lindbladian without constructing the full superoperator, whose matrix representation has dimensions \(D^2 \times D^2\), with \(D\) being the Hilbert-space dimension. As a result, explicitly building the Liouvillian quickly becomes computationally prohibitive as the system size increases.

We consider a periodically driven open quantum system whose stroboscopic dynamics is governed by the Lindblad master equation of period $T$,
\begin{equation}
    \partial_t \hat{\rho}(t) = \mathcal{L}(t)\,\hat{\rho}(t), 
    \qquad \mathcal{L}(t) = \mathcal{L}(t+T).
    \label{eq:A1_Arnoldi}
\end{equation}
The evolution over one period is formally given by the time-ordered exponential
\begin{equation}
    \hat{\rho}(T) = 
    \mathcal{T}\!\left[ \exp\!\left( \int_{0}^{T} \mathcal{L}(t')\,dt' \right) \right]\hat{\rho}(0)
     \equiv \mathcal{F}(T,0)\,\hat{\rho}(0),
    \label{eq:A2_Arnoldi}
\end{equation}
where $\mathcal{F}(T,0)$ denotes the {Floquet map} associated with one full period of evolution.

By periodicity, 
\begin{equation}
    \mathcal{F}(t,0) = [\mathcal{F}(T,0)]^{n}\,\mathcal{F}(t-nT,0),
    \label{eq:A3_Arnoldi}
\end{equation}
for integer $n$. Defining the Floquet Liouvillian $\mathcal{L}_F$ through
\begin{equation}
    \mathcal{F}(T,0) = \exp(\mathcal{L}_F T),
    \label{A4_Arnoldi}
\end{equation}
one obtains
\begin{equation}
    \mathcal{F}\,\hat{\rho}_j^F = \varphi_j^F \hat{\rho}_j^F 
    = e^{\lambda_j^F T}\hat{\rho}_j^F 
    \quad \iff \quad 
    \mathcal{L}_F \hat{\rho}_j^F = \lambda_j^F \hat{\rho}_j^F.
\end{equation}
The eigenvalues $\lambda^F_j$ form the Floquet-Liouvillian spectrum, from which the Liouvillian gap $g = -\max_{\alpha\neq 0}\mathrm{Re}\,\lambda^F_\alpha$ is extracted.

In practice, we generate the Krylov subspace
\begin{equation}
\begin{aligned}
    \mathcal{K}_n &= \{ \hat{\rho}(0), \hat{\rho}(T), \hat{\rho}(2T), \ldots, \hat{\rho}(nT) \} \\
                  &= \{ \hat{\rho}(0), \mathcal{F}\hat{\rho}(0), \mathcal{F}^2\hat{\rho}(0), \ldots, \mathcal{F}^n\hat{\rho}(0) \}.
\end{aligned}
\label{eq:Krylov}
\end{equation}
directly from time evolution under the Lindblad equation, without explicitly constructing $\mathcal{L}$ or $\mathcal{F}$. Applying the Arnoldi iteration to this subspace yields an effective Hessenberg matrix whose eigenvalues approximate those of $\mathcal{F}$, providing the leading relaxation rates of the Liouvillian spectrum. The key advantage of this method is that the resulting Hessenberg matrix has dimensions \( n \times n \), with \( n \ll D^2 \), thus drastically reducing the computational cost compared to handling the full Liouvillian superoperator.

We validated the implementation by explicitly constructing $\mathcal{L}$ for small system sizes ($L=4,5,6$) and comparing the resulting eigenvalues with those obtained from the Arnoldi--Lindblad method. The agreement was excellent, confirming the accuracy and efficiency of the approach.
%%%%%%%%%%%%%%%%% Biblio %%%%%%%%%%%%%%%%%%%%%%%%%
%\bibliography{refs}  
\providecommand{\noopsort}[1]{}\providecommand{\singleletter}[1]{#1}%
%
%%%%%%%%%%%%%%%%%%%%%%%%%%%%%
\end{document}